\documentclass[aps,prd,showpacs,twocolumn,superscriptaddress,floatfix]{revtex4}
\usepackage{graphicx}  
\usepackage{dcolumn}   
\usepackage{bm}        
\usepackage[english]{babel}
\usepackage{amsfonts,amsmath,amssymb,mathrsfs}
\usepackage{color}
\usepackage{times}

\newcommand{\eq}{\begin{equation}}
\newcommand{\eeq}{\end{equation}}
\newcommand{\be}{\begin{equation}}
\newcommand{\ee}{\end{equation}}
\newcommand{\bea}{\begin{eqnarray}}

\newcommand{\eea}{\end{eqnarray}}
\newcommand{\vta}[1]{\vert \boldsymbol{a}_{#1}        \vert}
\newcommand{\vtaf}  {\vert \boldsymbol{a}_{\rm fin}   \vert}
\newcommand{\vtl}   {\vert \boldsymbol{       {\ell}} \vert}
\newcommand{\bl}    {      \boldsymbol{\tilde {\ell}}      }

\begin{document}

\title{On the final spin from the coalescence of two black holes}

\author{Luciano Rezzolla}
\affiliation{Max-Planck-Institut f\"ur Gravitationsphysik,
Albert-Einstein-Institut, Potsdam-Golm, Germany}
\affiliation{Department of Physics and Astronomy, Louisiana State
University, Baton Rouge, LA, USA }
\author{Enrico Barausse}
\affiliation{SISSA, International School for Advanced Studies and
INFN, Via Beirut 2, 34014 Trieste, Italy}
\author{Ernst Nils Dorband}
\affiliation{Max-Planck-Institut f\"ur Gravitationsphysik,
Albert-Einstein-Institut, Potsdam-Golm, Germany}
\author{Denis Pollney}
\affiliation{Max-Planck-Institut f\"ur Gravitationsphysik,
Albert-Einstein-Institut, Potsdam-Golm, Germany}
\author{Christian Reisswig}
\affiliation{Max-Planck-Institut f\"ur Gravitationsphysik,
Albert-Einstein-Institut, Potsdam-Golm, Germany}
\author{Jennifer Seiler}
\affiliation{Max-Planck-Institut f\"ur Gravitationsphysik,
Albert-Einstein-Institut, Potsdam-Golm, Germany}
\author{Sascha Husa}
\affiliation{Max-Planck-Institut f\"ur Gravitationsphysik,
Albert-Einstein-Institut, Potsdam-Golm, Germany}

\date{\today}
\begin{abstract}
We provide a compact analytic formula to compute the spin of the black
hole produced by the coalescence of two black holes following a
quasi-circular inspiral. Without additional fits than those already
available for binaries with aligned or antialigned spins, but with a
minimal set of assumptions, we derive an expression that can model
generic initial spin configurations and mass ratios, thus covering all
of the 7-dimensional space of parameters. A comparison with
simulations already shows very accurate agreements with all of the
numerical data available to date, but we also suggest a number of ways
in which our predictions can be further improved.
 \end{abstract}

\maketitle

The evolution of black hole binary systems is one of the most
important problems for general relativity, and more recently for
astrophysics, as such systems enter the realm of observation.  Recent
advances in numerical relativity have made it possible to cover the
entire range of the inspiral process, from large separations at which
post-Newtonian (PN) calculations provide accurate orbital parameters,
through the highly relativistic merger, to ringdown. For many studies
of astrophysical interest, such as many-body studies of galactic
mergers, or heirarchical models of black-hole formation however, it is
impractical to carry out evolutions with the full Einstein, or even
post-Newtonian, equations. Fortunately, recent binary black-hole
evolutions in full general relativity have shown that certain physical
quantities can be estimated to good accuracy if the initial encounter
parameters are known. In particular, this paper develops a rather
simple and robust formula for determining the spin of the black-hole
remnant resulting from the merger of rather generic initial binary
configurations.

To appreciate the spirit of our approach it can be convenient to think
of the inspiral and merger of two black holes as a mechanism which
takes, as input, two black holes of initial masses $M_{1}$, $M_{2}$
and spin vectors $\boldsymbol{S}_{1}$, $\boldsymbol{S}_{2}$ and
produces, as output, a third black hole of mass $M_{\rm fin}$ and spin
$\boldsymbol{S}_{\rm fin}$. In conditions of particular astrophysical
interest, the inspiral takes place through quasi-circular orbits since
the eccentricity is removed quickly by the gravitational-radiation
reaction~\cite{Peters:1964}. Furthermore, at least for nonspinning
equal-mass black holes, the final spin does not depend on the value of
the eccentricity as long as it is not too
large~\cite{Hinder:2007qu}. The determination of $M_{\rm fin}$ and
$\boldsymbol{S}_{\rm fin}$ from the knowledge of $M_{1,2}$ and
$\boldsymbol{S}_{1,2}$, is of great importance in several fields. In
astrophysics, it provides information on the properties of isolated
stellar-mass black holes produced at the end of the evolution of a
binary system of massive stars. In cosmology, it can be used to model
the distribution of masses and spins of the supermassive black holes
produced through the merger of galaxies (see ref.~\cite{Berti2008} for
an interesting example). In addition, in gravitational-wave astronomy,
the a-priori knowledge of the final spin can help the detection of the
ringdown.  What makes this a difficult problem is clear: for binaries
in quasi-circular orbits the space of initial parameters for the final
spin has seven dimensions (\textit{i.e.}, the mass-ratio $q\equiv
M_2/M_1$ and the six components of the spin vectors). A number of
analytical approaches have been developed over the years to determine
the final spin, either exploiting the dynamics of
point-particles~\cite{Hughes:2002ei,Buonanno:07b} or the PN
approximation~\cite{Gergely:07}, or using more sophisticated
approaches such as the effective-one-body
approximation~\cite{Buonanno:06cd}. Ultimately, however, computing
$\boldsymbol{a}_{\rm fin} \equiv \boldsymbol{S}_{\rm fin}/M^2_{\rm
  fin}$ accurately requires the solution of the full Einstein
equations and thus the use of numerical-relativity
simulations. Several groups have investigated this problem over the
last couple of years~\cite{Campanelli:2006vp, Pollney:2007ss,
  Bruegmann:2007zj, Rezzolla-etal-2007,
  Marronetti07tbgs,Rezzolla-etal-2007b}.

While the recent possibility of measuring accurately the final spin
through numerical-relativity calculations represents an enormous
progress, the complete coverage of the full parameter space uniquely
through simulations is not a viable option. As a consequence, work has
been done to derive analytic expressions for the final spin which
would model the numerical-relativity data but also exploit as much
information as possible either from perturbative studies, or from the
symmetries of the system~\cite{Pollney:2007ss, Rezzolla-etal-2007,
  BoyleKesdenNissanke:07, BoyleKesden:07,
  Marronetti07tbgs,Rezzolla-etal-2007b}.  In this sense, these
approaches do not amount to a blind fitting of the
numerical-relativity data, but, rather, use the data to construct a
physically consistent and mathematically accurate modelling of the
final spin. Despite a concentrated effort in this direction, the
analytic expressions for the final spin could, at most, cover 3 of the
7 dimensions of the space of
parameters~\cite{Rezzolla-etal-2007b}. Here, we show that without
additional fits and with a minimal set of assumptions it is possible
to obtain the extension to the complete space of parameters and
reproduce all of the available numerical-relativity data. Although our
treatment is intrinsically approximate, we also suggest how it can be
improved.

Analytic fitting expressions for $\boldsymbol{a}_{\rm fin}$ have so
far been built using binaries having spins that are either
\textit{aligned} or \textit{antialigned} with the initial orbital
angular momentum. This is because in this case both the initial and
final spins can be projected in the direction of the orbital angular
momentum and it is possible to deal simply with the (pseudo)-scalar
quantities $a_1$, $a_2$ and $a_{\rm fin}$ ranging between $-1$ and
$+1$. If the black holes have \textit{equal mass} but \textit{unequal} spins
that are either \textit{parallel} or \textit{antiparallel}, then the
spin of the final black hole has been shown to be accurately described by the
simple analytic fit~\cite{Rezzolla-etal-2007}
\begin{equation}
\label{eqmass_uneqspin}
a_{\rm fin}(a_1,a_2)=p_0 + p_1 (a_1 + a_2) + p_2 (a_1 + a_2)^2\,,
\end{equation}
where $p_0 = 0.6883 \pm 0.0003$, $p_1 = 0.1530 \pm 0.0004$, and $p_2 =
-0.0088 \pm 0.0005$. When seen as a power series of the initial spins,
expression~\eqref{eqmass_uneqspin} suggests an interesting physical
interpretation. Its zeroth-order term, in fact, can be associated with
the (dimensionless) orbital angular momentum not radiated in
gravitational waves and amounting to $\sim 70\%$ of the final spin at
most. The first-order term, on the other hand, can be seen as the
contributions from the initial spins and from the spin-orbit coupling,
amounting to $\sim 30\%$ at most. Finally, the second-order term,
includes the spin-spin coupling, with a contribution to the final spin
which is of $\sim 4\%$ at most.

If the black holes have \textit{unequal mass} but spins that are
\textit{equal} and \textit{parallel}, the final spin is instead given
by the analytic fit~\cite{Rezzolla-etal-2007b}
\begin{eqnarray}
\label{eqspin_uneqmass}
&&a_{\rm fin}(a,\nu)=a+s_{4}a^2 \nu+s_{5}a
	\nu^2+t_{0} a\nu+ \nonumber \\
&& \hskip 1.75cm 2\sqrt{3}\nu+t_2\nu^2+t_{3}\nu^3\,,
\end{eqnarray}
where $\nu$ is the symmetric mass ratio $\nu \equiv
M_1M_2/(M_1+M_2)^2$, and where the coefficients take the values $s_4 =
-0.129 \pm 0.012$, $s_5 = -0.384 \pm 0.261$, $t_0 = -2.686 \pm 0.065$,
$t_2 = -3.454 \pm 0.132$, $t_3 = 2.353 \pm 0.548$. Although obtained
independently in~\cite{Rezzolla-etal-2007}
and~\cite{Rezzolla-etal-2007b}, expressions~\eqref{eqmass_uneqspin}
and~\eqref{eqspin_uneqmass} are compatible as can be seen by
considering~\eqref{eqspin_uneqmass} for equal-mass binaries ($\nu
=1/4$) and verifying that the following relations hold within the
computed error-bars
\begin{equation}
\label{relations}
p_0= \frac{\sqrt{3}}{2} + \frac{t_2}{16} + \frac{t_3}{64}\,,
\quad p_1 = \frac{1}{2} + \frac{s_5}{32} + 
	\frac{t_0}{8}\,, \quad p_2 = \frac{s_4}{16}.
\end{equation}

As long as the initial spins are aligned (or antialigned) with the
orbital angular momentum, expression~\eqref{eqspin_uneqmass} can be
extended to \textit{unequal-spin, unequal-mass} binaries through the
substitution
\begin{equation}
\label{substitution}
    a \ \to \ \tilde{a} \equiv \frac{a_1 + a_2 q^2}{1+q^2} \,.
\end{equation}
To obtain this result, it is sufficient to
consider~\eqref{eqmass_uneqspin} and~\eqref{eqspin_uneqmass} as
polynomial expressions of the generic quantity
\begin{equation}
    \tilde{a} \equiv a_{\rm tot} \frac{(1+q)^2}{1+q^2}\,.
\end{equation}
where $a_{\rm tot} \equiv (a_1 + a_2 q^2)/(1+q)^2$ is the total
dimensionless spin for generic aligned binaries. In this way,
expressions~\eqref{eqmass_uneqspin} and ~\eqref{eqspin_uneqmass} are
naturally compatible, since $\tilde{a} = (a_1+a_2)/2$ for equal-mass
unequal-spin binaries, and $\tilde{a} = a$ for unequal-mass equal-spin
binaries. Furthermore, the extreme mass-ratio limit (EMRL) of
expression~\eqref{eqspin_uneqmass} with the
substitution~\eqref{substitution} yields the expected result: $a_{\rm
  fin}(a_1, a_2, \nu=0) = a_1$.

As already commented above, the predictions of
expressions~\eqref{eqspin_uneqmass} and~\eqref{substitution} cover 3
of the 7 dimensions of the space of parameters for binaries in
quasi-circular orbits; we next show how to to cover the remaining 4
dimensions and derive an analytic expression for the dimensionless
spin \textit{vector} $\boldsymbol{a}_{\rm fin}$ of the black hole
produced by the coalescence of two generic black holes in terms of the
mass ratio $q$ and of the initial dimensionless spin vectors
$\boldsymbol{a}_{1,2}$. To make the problem tractable analytically, 4
assumptions are needed. While some of these are very natural, others
can be relaxed if additional accuracy in the estimate of
$\boldsymbol{a}_{\rm fin}$ is necessary. It should be noted, however,
that removing any of these assumptions inevitably complicates the
picture, introducing additional dimensions, such as the initial
separation in the binary or the radiated mass, in the space of
parameters.

As a result, in the simplest and yet accurate description the required
assumptions are as follows:

\smallskip
\textit{(i) The mass radiated to gravitational waves $M_{\rm rad}$ can
  be neglected} \textit{i.e.}, $M_{\rm fin} = M \equiv M_1 + M_2$.  We
note that $M_{\rm rad}/M = 1-M_{\rm fin}/M \approx 5-7\times 10^{-2}$
for most of the binaries evolved numerically. The same assumption was
applied in the analyses
of~\cite{Rezzolla-etal-2007,Rezzolla-etal-2007b}, as well as
in~\cite{Buonanno:07b}. Relaxing this assumption would introduce a
dependence on $M_{\rm fin}$ which can only be measured through a
numerical simulation.

\smallskip
\textit{(ii) At a sufficiently large but finite initial separation the
  final spin vector $\boldsymbol{S}_{\rm fin}$ can be well
  approximated as the sum of the two initial spin vectors and of a
  third vector $\boldsymbol{\tilde{\ell}}$}
\begin{equation}
\label{assumption_1}
\boldsymbol{S}_{\rm fin}=\boldsymbol{S}_1+
\boldsymbol{S}_2+\boldsymbol{\tilde{\ell}}\,,
\end{equation}
Differently from refs.~\cite{Hughes:2002ei} and~\cite{Buonanno:07b},
where a definition similar to \eqref{assumption_1} was also
introduced, here we will constrain $\boldsymbol{\tilde{\ell}}$ by
exploiting the results of numerical-relativity calculations rather
than by relating it to the orbital angular momentum of a test particle
at the innermost stable circular orbit (ISCO). When viewed as
expressing the conservation of the total angular momentum,
eq.~\eqref{assumption_1} also defines the vector $\bl$ as the
difference between the orbital angular momentum when the binary is
widely separated $\boldsymbol{L}$, and the angular momentum radiated
until the merger $\boldsymbol{J}_{\rm rad}$, \textit{i.e.}, $\bl =
\boldsymbol{L} - \boldsymbol{J}_{\rm rad}$. 

\smallskip \textit{(iii) The vector $\boldsymbol{\tilde{\ell}}$ is
  parallel to $\boldsymbol{L}$}. This assumption is correct when
$\boldsymbol{S}_1=-\boldsymbol{S}_2$ and $q=1$ [this can be seen from
  the PN equations at 2.5 order], or by equatorial
symmetry when the spins are aligned with $\boldsymbol{L}$ or when
$\boldsymbol{S}_1=\boldsymbol{S}_2=0$ (also these cases can be seen
from the PN equations). For more general configurations one expects
that $\bl$ will also have a component orthogonal to $\boldsymbol{L}$
as a result, for instance, of spin-orbit or spin-spin couplings, which
will produce in general a precession of $\bl$. In practice, the
component of $\bl$ orthogonal to $\boldsymbol{L}$ will correspond to
the angular momentum $\boldsymbol{J}^{\perp}_{\rm rad}$ radiated in a
plane orthogonal to $\boldsymbol{L}$, with a resulting error in the
estimate of $\vert \bl \vert$ which is $\sim \vert
\boldsymbol{J}^{\perp}_{\rm rad} \vert^2 / \vert
\boldsymbol{\tilde{\ell}} \vert^2\sim \vert
\boldsymbol{J}^{\perp}_{\rm rad} \vert^2/(2 \sqrt{3} M_1
M_2)^2$\footnote{Assumption \textit{(iii)} can be equivalently
  interpreted as enforcing that the component of the final spin
  $\boldsymbol{S}_{\rm fin}$ in the orbital plane equals the one of the
  total initial spin $\boldsymbol{S}_1+ \boldsymbol{S}_2$ in that
  plane.}. Although these errors are small in all the configurations
that we have analysed, they may be larger in general
configurations. Measuring $\boldsymbol{J}^{\perp}_{\rm rad}$ via
numerical-relativity simulations, or estimating it via high-order PN
equations, is an obvious way to improve our approach. A similar
assumption was also made in ref.~\cite{Buonanno:07b}.

\smallskip
\textit{(iv) When the initial spin vectors are equal and opposite
  ($\boldsymbol{S}_{1}=-\boldsymbol{S}_{2}$) and the masses are equal
  ($q=1$), the spin of the final black hole is the same as for the
  nonspinning binaries}. Stated differently, equal-mass binaries with
equal and opposite-spins behave as nonspinning binaries, at least when
it comes down to the properties of the final black hole. While this
result cannot be derived from first principles, it reflects the
expectation that if the spins are the same and opposite, their
contributions to the final spin cancel for equal-mass
binaries. Besides being physically reasonable, this expectation is met
by all of the simulations performed to date, both for spins aligned
with $\boldsymbol{L}$~\cite{Rezzolla-etal-2007,Rezzolla-etal-2007b}
and orthogonal to $\boldsymbol{L}$~\cite{Bruegmann:2007zj}. In
addition, this expectation is met by the leading-order contributions
to the spin-orbit and spin-spin point-particle Hamiltonians and
spin-induced radiation flux~\cite{Barker1970,Buonanno:06cd}.  A
similar assumption is also made, although not explicitly, in
ref.~\cite{Buonanno:07b} which, for $\boldsymbol{S}_{\rm tot} = 0$,
predicts $\iota=0$ and $\vtaf = L_{\rm orb}(\iota=0,\vtaf)/M
=\,$const. [\textit{cf.}~eqs. (12)--(13) in ref.~\cite{Buonanno:07b}].

\smallskip

Using these assumptions we can now derive the analytic expression for
the final spin. We start by expressing the vector
relation~\eqref{assumption_1} as
\begin{equation}
\label{assumption_1_bis}
\boldsymbol{a}_{\rm
fin}=\frac{1}{(1+q)^2}
\left(\boldsymbol{a}_1+\boldsymbol{a}_2q^2 + 
\boldsymbol{{\ell}} q \right)\,,
\end{equation}
where $\boldsymbol{a}_{\rm fin}= \boldsymbol{S}_{\rm fin}/M^2$
[\textit{cf.} assumption \textit{(i)}], $\boldsymbol{{\ell}} \equiv
\boldsymbol{\tilde{\ell}}/(M_1 M_2)$,
\hbox{$\boldsymbol{a}_{1,2}\equiv \boldsymbol{S}_{1,2}/M^2_{1,2}$},
and its norm is then given by
\begin{align}
\label{eq:general}
&\hskip -0.5cm \quad \vert \boldsymbol{a}_{\rm fin}\vert=
\frac{1}{(1+q)^2}\Big[ \vta{1}^2 + \vta{2}^2 q^4+
 2 {\vert \boldsymbol{a}_2\vert}{\vert 
\boldsymbol{a}_1\vert} q^2 \cos \alpha +
\nonumber\\ 
& \hskip 1.cm 
2\left(
     {\vert \boldsymbol{a}_1\vert}\cos \beta +
     {\vert \boldsymbol{a}_2\vert} q^2  \cos \gamma
\right) {\vert \boldsymbol{{\ell}} \vert}{q}
+ \vert \boldsymbol{{\ell}}\vert^2 q^2
\Big]^{1/2}\,,
\end{align}
where the three (cosine) angles $\alpha, \beta$ and $\gamma$ are 
defined by
\begin{equation}
\label{cosines}
\cos \alpha \equiv
{\boldsymbol{\hat{a}}_1\cdot\boldsymbol{\hat{a}}_2}
\,,
\hskip 0.3cm
\cos \beta \equiv
 \boldsymbol{\hat a}_1\cdot\boldsymbol{\hat{{\ell}}}\,,
\hskip 0.3cm
\cos \gamma \equiv
\boldsymbol{\hat{a}}_2\cdot\boldsymbol{\hat{{\ell}}}\,.
\end{equation}
Because $\boldsymbol{a}_{1,2} \parallel \boldsymbol{S}_{1,2}$ and 
$\boldsymbol{{\ell}} \parallel \boldsymbol{L}$ [\textit{cf.} assumption \textit{(iii)}], the angles $\alpha,
\beta$ and $\gamma$ are also those between the initial spin vectors 
and the initial orbital angular momentum, so that it is possible to replace
$\boldsymbol{\hat{a}}_{1,2}$ with $\boldsymbol{\hat{S}}_{1,2}$ and $\boldsymbol{\hat{\ell}}$ with
$\boldsymbol{\hat{L}}$ in~\eqref{cosines}. Note that $\alpha, \beta$ and $\gamma$ are well-defined if the initial
separation of the two black holes is sufficiently large [\textit{cf.}
  assumption \textit{(ii)}] and that the error introduced by assumption \textit{(iii)} in the measure of 
$\cos\alpha, \cos\beta$ and $\cos\gamma$ is also of the order of $\vert
\boldsymbol{J}^{\perp}_{\rm rad} \vert / \vert
\boldsymbol{\tilde{\ell}} \vert$.

The angle $\theta_{\rm fin}$ between the final spin vector and the
initial orbital angular momentum can be easily calculated from
$\vtaf$. Because of assumption \textit{(iii)}, the component of the
final spin in the direction of $\boldsymbol{L}$ is [\textit{cf.}
eq.~\eqref{assumption_1_bis}]
\begin{equation}
\label{eq:a_z}
a_{\rm fin}^{\parallel}\equiv 
\boldsymbol{a}_{\rm fin}\cdot \boldsymbol{\hat{{\ell}}}
= \frac{\vta{1} \cos\beta + 
  \vta{2}q^2 \cos\gamma  + \vtl q}{(1+q)^2}\,,
\end{equation}
so that $\cos\theta_{\rm fin}={a_{\rm fin}^{\parallel}}/{ \vert
\boldsymbol{a}_{\rm fin}\vert}$, and the component orthogonal to the
initial orbital angular momentum is $a_{\rm fin}^{\perp} = \vtaf
\sin\theta_{\rm fin}$.

In essence, therefore, our approach consists of considering the
dimensionless spin vector of the final black hole as the sum of the two
initial spins and of a third vector parallel to the initial orbital
angular momentum when the binaries are widely separated. Implicit in
the assumptions made, and in the logic of mapping an initial-state of
the binary into a final one, is the expectation that the length of
this vector is an intrinsic ``property'' of the binary, depending on
the initial spin vectors and mass ratio, but not on the initial
separation. This is indeed a consequence of assumption \textit{(ii)}:
because the vector $\boldsymbol{\tilde{\ell}} $ measures the orbital
angular momentum that cannot be radiated, it can be thought of as the
angular momentum of the binary at the ``effective'' ISCO and, as such,
it cannot be dependent on the initial separation. 

A very important consequence of our assumptions is that
$\boldsymbol{a}_{\rm fin}$ for a black-hole binary is already fully
determined by the set of coefficients $s_4, s_5, t_0, t_2, t_3$
computed to derive expression~\eqref{eqspin_uneqmass}. The latter, in
fact, is simply the final spin for a special set of values for the
cosine angles; since the fitting coefficients are constant, they must
hold also for generic binaries.

In view of this, all that is needed is to measure $\vert \boldsymbol{\ell}
\vert$ in terms of the fitting coefficients computed in
refs.~\cite{Rezzolla-etal-2007,Rezzolla-etal-2007b}. This can be done
by matching expression~\eqref{eq:a_z} with~\eqref{eqspin_uneqmass}
[with the condition~\eqref{substitution}] for parallel and aligned
spins ($\alpha=\beta=\gamma=0$), for parallel and antialigned spins
($\alpha=0$, $\beta=\gamma=\pi$), and for antiparallel spins which are
aligned or antialigned ($\alpha=\beta=\pi$, $\gamma=0$ or
$\alpha=\gamma=\pi$, $\beta=0$). This matching is not unique, but the
degeneracy can be broken by exploiting assumption \textit{(iv)} and by
requiring that $\vtl$ depends linearly on $\cos\alpha$, $\cos\beta$
and $\cos\gamma$. We therefore obtain
\begin{eqnarray}
\label{eq:L}
&&
\hskip -0.5cm
\vtl
 = 
\frac{s_4}{(1+q^2)^2} \left(\vta{1}^2 + \vta{2}^2 q^4 
	+ 2 \vta{1} \vta{2} q^2 \cos\alpha\right) + 
\nonumber \\
&& \hskip 0.5cm
\left(\frac{s_5 \nu + t_0 + 2}{1+q^2}\right)
	\left(\vta{1}\cos\beta + \vta{2} q^2 \cos\gamma\right) +
\nonumber \\
&& \hskip 0.5cm
	2 \sqrt{3}+ t_2 \nu + t_3 \nu^2 \,.
\end{eqnarray}
\begin{figure*}[tbp!]
\centerline{
\resizebox{8cm}{!}{\includegraphics[angle=-0]{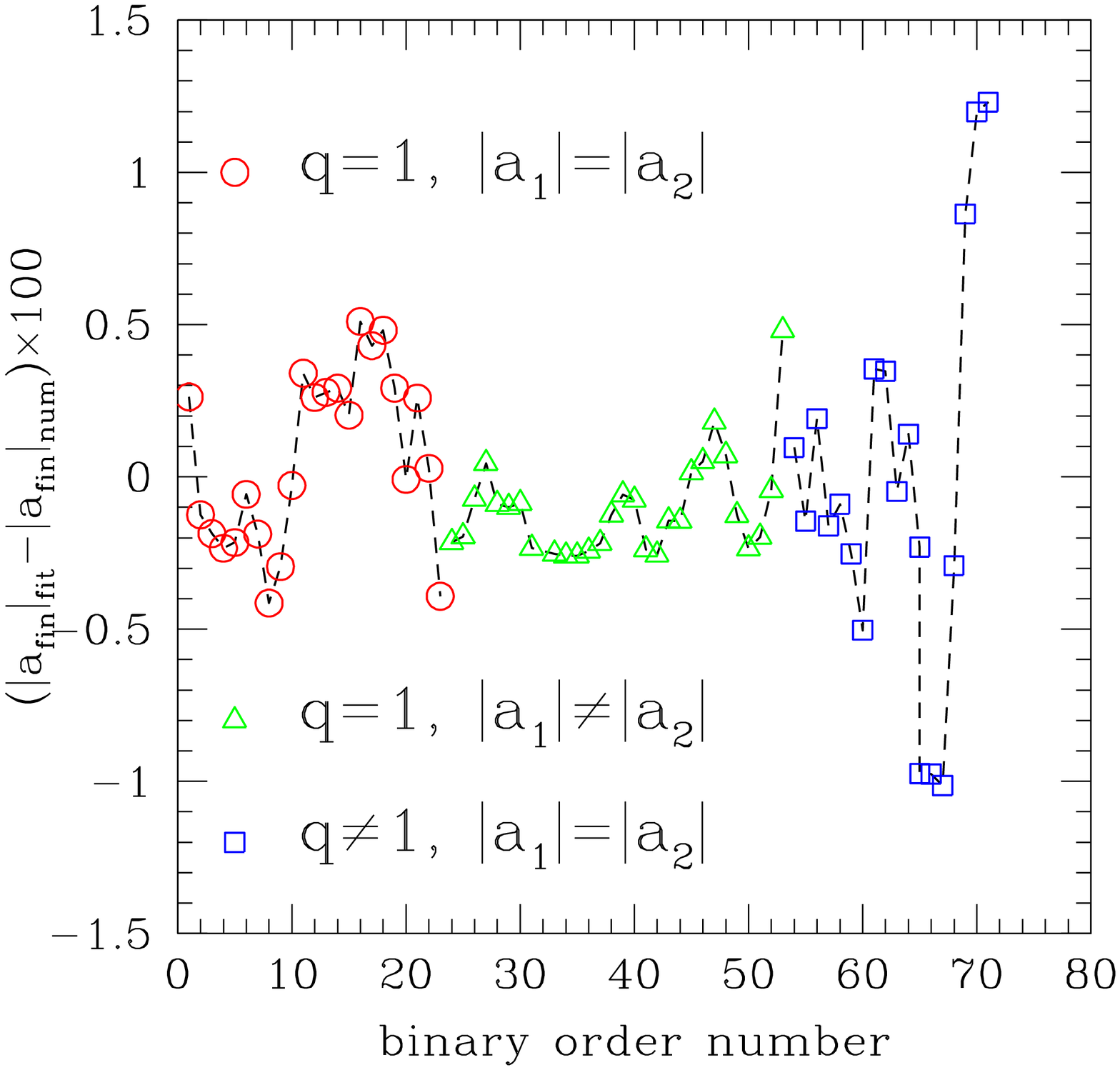}}
\hskip 1.0cm
\resizebox{8cm}{!}{\includegraphics[angle=-0]{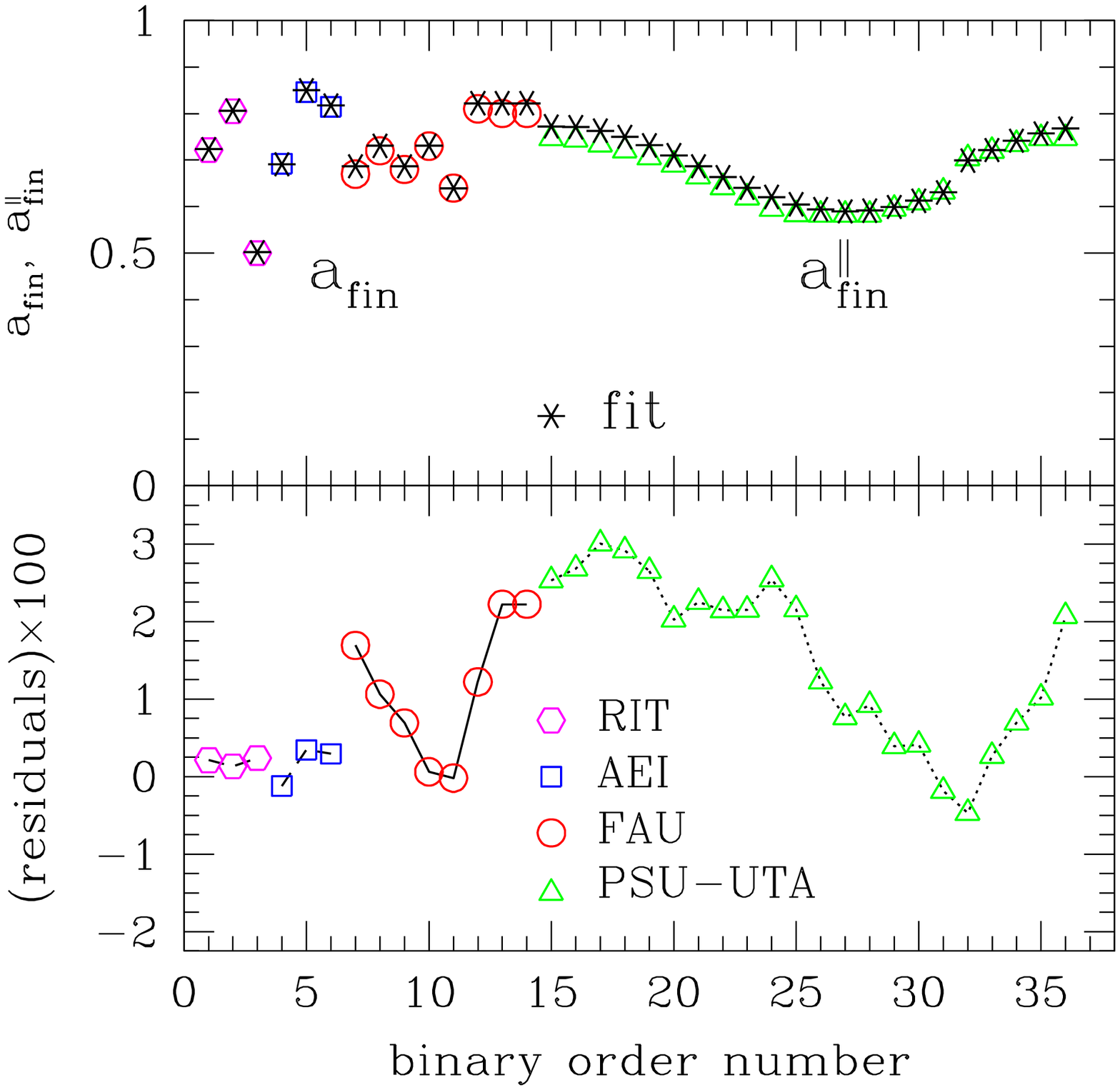}}
}
\vskip -0.5cm
\caption{\textit{Left panel:} Rescaled residual for aligned
  binaries. The circles refer to equal-mass, equal-spin binaries
  presented in refs.~\cite{Rezzolla-etal-2007, Marronetti07tbgs,
    Berti:2007snb, Berti:2007sb, Buonanno:2007ft,
    Rezzolla-etal-2007b}, triangles to equal-mass, unequal-spin
  binaries presented in ref.~\cite{Rezzolla-etal-2007, Berti:2007sb},
  and squares to unequal-mass, equal-spin binaries presented in
  refs.~\cite{Berti:2007snb,Buonanno:2007ft,Rezzolla-etal-2007b,Berti:2007sb}.
  Here and in the right panel the ``binary order number'' is just a
  dummy index labelling the different configurations.  \textit{Right
    panel:} The top part reports with asterisks the final spin computed for
  misaligned binaries. Hexagons refer to data from
  ref.~\cite{Campanelli:2006vp} (labelled ``RIT''), squares to the
  data Table~\ref{tableone} (labelled ``AEI''), circles to data from
  ref.~\cite{Tichy:2007gso} (labelled ``FAU''), and triangles to data
  from ref.~\cite{Herrmann:2007ex} (labelled ``PSU-UTA''). Note that
  these latter data points refer to the aligned component $a_{\rm
    fin}^{\parallel}$ since this is the only component available from
  ref.~\cite{Herrmann:2007ex}. The bottom part of this panel shows
  instead the rescaled residuals for these misaligned binaries.}
\label{fig:align_res}
\vskip -0.5cm
\end{figure*}

We now consider some limits of expressions~\eqref{eq:general}
and~\eqref{eq:L}. First of all, when $q\to0$,~\eqref{eq:general}
and~\eqref{eq:L} yield the correct EMRL, \textit{i.e.}, $\vtaf =
\vta{1}$. Secondly, for equal-mass binaries having spins that are
equal and antiparallel,~\eqref{eq:general} and~\eqref{eq:L} reduce to
\begin{equation}
\label{2nd_check_bis}
\vtaf = \frac{\vtl}{4} = \frac{\sqrt{3}}{2} + \frac{t_2}{16} +
	\frac{t_3}{64} = p_0 \simeq 0.687 \,.
\end{equation}

This result allows us now to qualify more precisely a comment made
before: because for equal-mass black holes which are either nonspinning or
have equal and opposite spins, the vector $\vtl$ does not depend on
the initial spins, expression~\eqref{2nd_check_bis} states that $\vert
\boldsymbol{{\ell}}\vert M_{\rm fin}^2/4=\vert
\boldsymbol{{\ell}}\vert M^2/4=\vert \boldsymbol{{\ell}}\vert M_1 M_2$ is, for such systems, the orbital
angular momentum at the effective ISCO. We can take this a step
further and conjecture that $\vert \boldsymbol{{\ell}}\vert M_1 M_2
=\vert \boldsymbol{\tilde{\ell}}\vert$ is the series expansion of the
dimensionless orbital angular momentum at the ISCO also for \textit{unequal-mass}
binaries which are either nonspinning or with equal and opposite
spins. The zeroth-order term of this series (namely, the term
$2\sqrt{3} M_1 M_2$) is exactly the one predicted from the EMRL.  We
note that although numerical simulations do not reveal the presence of
an ISCO, the concept of an effective ISCO can nevertheless be useful
for the construction of gravitational-wave
templates~\cite{Ajith:2007kx,Hanna2008}.

Finally, we consider the case of equal, parallel and
aligned/antialigned spins ($\vta{2}=\vta{1}$, $\alpha=0$,
$\beta=\gamma=0,\,\pi$), for which expressions~\eqref{eq:a_z}
and~\eqref{eq:L} become
\begin{eqnarray}
\label{3rd_check}
a_{\rm fin} &=& \vta{1}\cos\beta \left[ 1 + \nu
(s_4\vta{1}\cos\beta + t_0 + s_5 \nu )\right] + \nonumber \\
&& \hskip 2.5cm	
\nu(
2 \sqrt{3}+ t_2 \nu + t_3 \nu^2 )\,,
\end{eqnarray}
where $\cos\beta = \pm 1$ for aligned/antialigned spins. As expected,
expression~\eqref{3rd_check} coincides with~\eqref{eqspin_uneqmass}
when $\vta{1}\cos\beta=a$ and with~\eqref{eqmass_uneqspin} [through
the coefficients~\eqref{relations}] when $q=1$ and $2\vta{1}\cos\beta
= a_1 + a_2$.  Similarly, \eqref{eq:a_z} and \eqref{eq:L} reduce
to~\eqref{eqspin_uneqmass} for equal, antiparallel and
aligned/antialigned spins ($\vta{2}=\vta{1}$, $\alpha=0$, $\beta=0,
\gamma=\pi$, or $\beta=\pi, \gamma=0$).
\begin{table}
\begin{ruledtabular}
\begin{tabular}{lccccccccc}
~                                       &
\multicolumn{1}{c}
{$a^x_1$}		                &
{$a^y_1$}		                &
{$a^z_1$}		                &
{$a^x_2$}		                &
{$a^y_2$}		                &
{$a^z_2$}		                &
{$\nu$}                                 &
{$\vtaf$}                 		&
{$\theta_{\rm fin}(^\circ)$}
\\
\hline
$~$       & 0.151 & 0.000 & -0.563 & 0.000 & 0.000 & 0.583 & 0.250 & 0.692 & 2.29 \\  
$~$       & 0.151 & 0.000 &  0.564 & 0.000 & 0.151 & 0.564 & 0.250 & 0.846 & 3.97 \\  
$~$       & 0.413 & 0.000 &  0.413 & 0.000 & 0.413 & 0.413 & 0.250 & 0.815 & 7.86 \\  
\end{tabular} 
\end{ruledtabular}
\vskip -0.25cm
\caption{\label{tableone}Initial parameters of the new misaligned AEI
binaries.}
\vskip -0.6cm
\end{table}

\begin{figure}[tbp!]
\centerline{
\resizebox{8cm}{!}{\includegraphics[angle=-0]{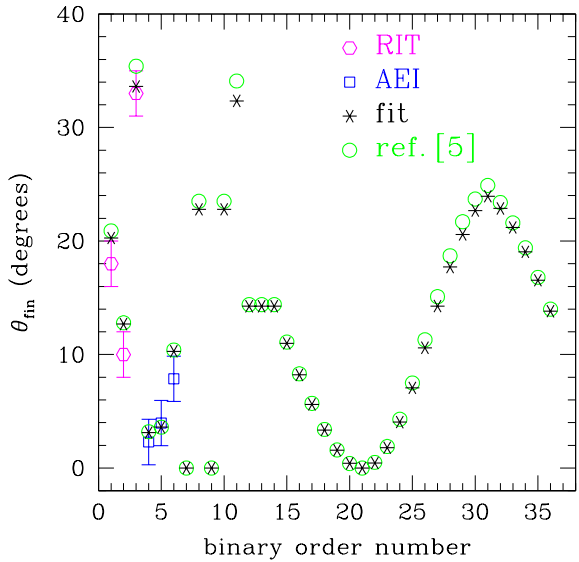}}
}
\vskip -0.5cm
\caption{Using the same data (and convention for the symbols) as in
  the right panel of Fig.~\ref{fig:align_res}, we here report the
  angle between the final spin vector and the initial orbital angular
  momentum $\theta_{\rm fin}$. Shown instead with asterisks and
  circles are the values predicted for the numerical data (as taken
  from refs.~\cite{Campanelli:2006vp,Tichy:2007gso,Herrmann:2007ex}
  and from Table~\ref{tableone}) by our analytic fit (asterisks) and
  by the point-particle approach suggested in ref.~\cite{Buonanno:07b}
  (circles).}
\label{fig:misalign}
\vskip -0.5cm
\end{figure}

The only way to assess the validity of expressions~\eqref{eq:general}
and~\eqref{eq:L} is to compare their predictions with the
numerical-relativity data. This is done in Figs.~\ref{fig:align_res}
and~\ref{fig:misalign}, which collect all of the published data,
together with the three additional binaries computed with the
\texttt{CCATIE} code~\cite{Pollney:2007ss} and reported in Table~\ref{tableone}. In these
plots, the ``binary order number'' is just a dummy index labelling the
different configurations. The left panel of Fig.~\ref{fig:align_res},
in particular, shows the rescaled residual, \textit{i.e.},
$(\vtaf_{\rm fit} - \vtaf_{\rm num.})\times 100$, for aligned
binaries. The plot shows the numerical-relativity data with circles
referring to equal-mass, equal-spin binaries from
refs.~\cite{Rezzolla-etal-2007, Marronetti07tbgs, Berti:2007snb,
  Berti:2007sb, Buonanno:2007ft, Rezzolla-etal-2007b}, triangles to
equal-mass, unequal-spin binaries from refs.~\cite{Rezzolla-etal-2007,
  Berti:2007sb}, and squares to unequal-mass, equal-spin binaries from
refs.~\cite{Berti:2007snb,Buonanno:2007ft,Rezzolla-etal-2007b,Berti:2007sb}. Although
the data is from simulations with different truncation errors, the
residuals are all very small and with a scatter of $\sim 1\%$.

A more stringent test is shown in the right panel of
Fig.~\ref{fig:align_res}, which refers to misaligned binaries. In the
top part, hexagons indicate the numerical values for $\vtaf$ from
ref.~\cite{Campanelli:2006vp}, squares the ones in
Table~\ref{tableone}, circles those from ref.~\cite{Tichy:2007gso} and
triangles those from ref.~\cite{Herrmann:2007ex}; note that these
latter data points refer to the aligned component $a_{\rm
  fin}^{\parallel}$ since this is the only component available from
ref.~\cite{Herrmann:2007ex}. The agreement is again very good, with
errors of a couple of percent (see bottom part of the same panel),
even if the binaries are generic and for some the initial and final
spins differ by almost $180^\circ$~\cite{Campanelli:2006vp}.

Finally, Fig.~\ref{fig:misalign} reports the angle between the final
spin vector and the initial orbital angular momentum $\theta_{\rm
  fin}$ using the same data (and convention for the symbols) as in the
right panel of Fig.~\ref{fig:align_res}. Measuring the final angle
accurately is not trivial, particularly due to the fact that the
numerical evolutions start at a finite separation which does not
account for earlier evolution of the orbital angular momentum
vector. The values reported in~\cite{Campanelli:2006vp} (and the
relative error-bars) are shown with hexagons, while the squares refer
to the binaries in Table~\ref{tableone}, and have been computed using
a new approach for the calculation of the Ricci scalar on the apparent
horizon~\cite{Jasiulek:2008dt}. Shown with asterisks and circles are
instead the values predicted for the numerical data (as taken from
refs.~\cite{Campanelli:2006vp,Tichy:2007gso,Herrmann:2007ex} and from
Table~\ref{tableone}) by our analytic fit (asterisks) and by the
point-particle approach suggested in ref.~\cite{Buonanno:07b}
(circles).

Clearly, when a comparison with numerical data is possible, the
estimates of our fit are in reasonable agreement with the data and
yield residuals in the final angle (\textit{i.e.}, $(\theta_{\rm
  fin})_{\rm fit} - (\theta_{\rm fin})_{\rm num.}$) which are
generally smaller than those obtained with the point-particle approach
of ref.~\cite{Buonanno:07b}. However, for two of the three binaries
from ref.~\cite{Campanelli:2006vp} the estimates are slightly outside
the error-bars. Note that the reported angles are relative to the
orbital plane at a small initial binary-separation, and thus are
likely to be underestimates as they do not take into account the
evolution from asymptotic distances; work is in progress to clarify
this. When the comparison with the numerical data is not possible
because $\theta_{\rm fin}$ is not reported (as for the data in
ref.~\cite{Herrmann:2007ex}), our approach and the one in
ref.~\cite{Buonanno:07b} yield very similar estimates.

In summary: we have considered the spin vector of the black hole produced by a
black-hole binary merger as the sum of the two initial spins and of a third
vector, parallel to the initial orbital angular momentum, whose norm
depends only on the initial spin vectors and mass ratio, and measures
the orbital angular momentum not radiated. Without additional fits
than those already available to model aligned/antialigned binaries, we
have measured the unknown vector and derived a formula that accounts
therefore for all of the 7 parameters describing a black-hole binary
inspiralling in quasi-circular orbits. 
The equations~\eqref{eq:general} and~\eqref{eq:L}, encapsulate the
near-zone physics to provide a convenient, but also robust and
accurate over a wide range of parameters, determination of the
merger product of rather generic black-hole binaries.

Testing the formula against all
of the available numerical data has revealed differences between the
predicted and the simulated values of a few percent at most. Our
approach is intrinsically approximate and it has been validated on a
small set of configurations, but it can be improved, for instance: by
reducing the $\chi^2$ of the fitting coefficients as new simulations
are carried out; by using fitting functions that are of higher-order
than those in expressions~\eqref{eqmass_uneqspin}
and~\eqref{eqspin_uneqmass}; by estimating
$\boldsymbol{J}^{\perp}_{\rm rad}$ through PN expressions or by
measuring it via numerical simulations.

\medskip
\noindent It is a pleasure to thank Peter Diener, Michael Jasiulek and
Erik Schnetter for valuable discussions. EB gratefully acknowledges
the hospitality of the AEI, where part of this work was carried
out. The computations were performed on the clusters Belladonna and
Damiana at the AEI.

\bibliography{published_version.bbl}

\end{document}